\documentclass[12pt]{iopart}

\usepackage{graphicx}
\begin{document}

\title[The Influence of Laser Focusing Conditions on DLA]{The Influence of Laser Focusing Conditions on the Direct Laser Acceleration of Electrons}

\author{H. Tang$^1$, K. Tangtartharakul$^2$, R.~Babjak$^{3,4}$, I-L.~Yeh$^5$, F.~Albert$^6$, H.~Chen$^6$, P.~T.~Campbell$^1$, Y.~Ma$^1$, P.~M.~Nilson$^7$, B.~K.~Russell$^1$, J.~L.~Shaw$^7$, A.~G.~R.~Thomas$^1$, M.~Vranic$^3$, A.~V.~Arefiev$^2$ and L.~Willingale$^1$}

\address{$^1$ G\'{e}rard Mourou Center for Ultrafast Optical Science, University of Michigan, 2200 Bonisteel Boulevard, Ann Arbor, Michigan 48109, USA}

\address{$^2$ Department of Mechanical and Aerospace Engineering, University of California at San Diego, La Jolla, California 92093, USA}

\address{$^3$ GoLP/Instituto de Plasmas e Fusão Nuclear, Instituto Superior Técnico, Universidade de Lisboa, Lisbon, 1049-001, Portugal}

\address{$^4$ Institute of Plasma Physics, Czech Academy of Sciences, Za Slovankou 1782/3, 182 00 Praha 8, Czechia}

\address{$^5$ Department of Physics, University of California at San Diego, La Jolla, California 92093, USA}

\address{$^6$ Lawrence Livermore National Laboratory, Livermore, California 94550, USA}

\address{$^7$ Laboratory for Laser Energetics, University of Rochester, Rochester, New York 14623, USA}

\ead{tanghm@umich.edu}

\vspace{10pt}
\begin{indented}
\item[]\today 
\end{indented}

\begin{abstract}
Direct Laser Acceleration (DLA) of electrons during a high-energy, picosecond laser interaction with an underdense plasma has been demonstrated to be substantially enhanced by controlling the laser focusing geometry.
Experiments using the OMEGA EP facility measured electrons accelerated to maximum energies exceeding 120 times the ponderomotive energy under certain laser focusing, pulse energy, and plasma density conditions.
Two-dimensional particle-in-cell simulations show that the laser focusing conditions alter the laser field evolution, channel fields generation, and electron oscillation, all of which contribute to the final electron energies. 
The optimal laser focusing condition occurs when the transverse oscillation amplitude of the accelerated electron in the channel fields matches the laser beam width, resulting in efficient energy gain.
Through this observation, a simple model was developed to calculate the optimal laser focal spot size in more general conditions and is validated by experimental data.
\end{abstract}

%
\vspace{2pc}

\section{Introduction}
The rapid advancement of ultra-high power laser facilities is enabling the realization of high energy density physics experiments and exploration of next-generation plasma-based accelerators \cite{strickland1985, mourou1997}. 
Direct laser acceleration (DLA) is one of the mechanisms for producing high-flux high-energy electrons \cite{pukhov1999, Gahn1999, meyer1999}. 
DLA has many potential applications in accelerating secondary ions/neutrons/positrons \cite{willingale2006, Pomerantz2014, Vranic2018, martinez2023creation}, providing bright directional x-rays \cite{Kneip2008}, material detection \cite{Ronga2021} and radiotherapy treatment\cite{Morimoto2018}. 
Understanding the energy transfer mechanism from the driving laser pulse to the plasma electrons is essential for these secondary processes.

In DLA process, the accelerated electrons gain energy directly from the laser. 
Consider a laser pulse with a peak normalized field strength $a_0= e E_0/(m_ec\omega_0)>1$, where $\omega_0$ is the laser frequency. 
The electric field of the laser oscillates the electron in the transverse direction and the $(\textbf{v}\times \textbf{B})$ force converts transverse momentum to longitudinal momentum. 
Previous studies depicted this process in a preformed channel \cite{arefiev2016}.
The channel fields dramatically increase electron energy gain by reducing the negative role of dephasing to keep electrons in phase with laser wave over an extended distance.
Simulations performed by K.~Tangtartharakul \textit{et al.} show that increasing the focal spot size and matched channel width can further enhance energy coupling efficiency because the negative work done by the longitudinal laser field component is reduced \cite{Kavin2023}.
Counter-intuitively, raising the laser peak intensity by focusing the beam to a smaller spot does not lead to higher electron energy.
R.~Babjak \textit{et al.} found an analytical prediction for the optimal focal spot, which approximately matches the electron transverse resonant amplitude, for highly relativistic intensity laser pulses 
over a range of plasma densities \cite{Babjak2023}.
Laser wakefield acceleration is another regime where focusing geometry has been considered an important parameter. 
A laser with a spot size larger than the plasma wavelength can maintain high intensity and be self-guided in plasma, allowing the acceleration of monoenergetic electron bunches \cite{thomas2007}.
However, we currently lack direct experimental evidence showing the effect of laser focusing geometries on DLA.

In this paper, the DLA electron energy dependency on laser focusing geometry is investigated using experiments and numerical modeling.
The experiments were performed on the OMEGA EP laser facility and demonstrated that the electron energy can be notably changed by varying the laser focal spot size.
This observation is strongly supported by complementary 2D particle-in-cell (PIC) OSIRIS simulations \cite{Osiris, Hemker2000}, which reveal how the evolution of the laser field and the generation of localized fields in plasma affect the accelerated electron motion inside the laser channel.
Particle tracking is performed to study the electron trajectories.
The relative energy contributions from the laser transverse and longitudinal fields are distinguished, as well as the detrimental effect of the sheath field formed at the rear of the target.
The sheath field strength increases with laser beam size, making its effect particularly notable for electrons accelerated by large beams.
We find the optimal focusing geometry can be analytically predicted by equating the electron transverse displacement and laser beam width.
This method is shown to fit the experimental results over a variety of plasma densities and laser parameters.

\section{Experimental setup}
\label{Exp_section}
\begin{figure}
		\begin{center}
		\includegraphics[width=.8\linewidth]{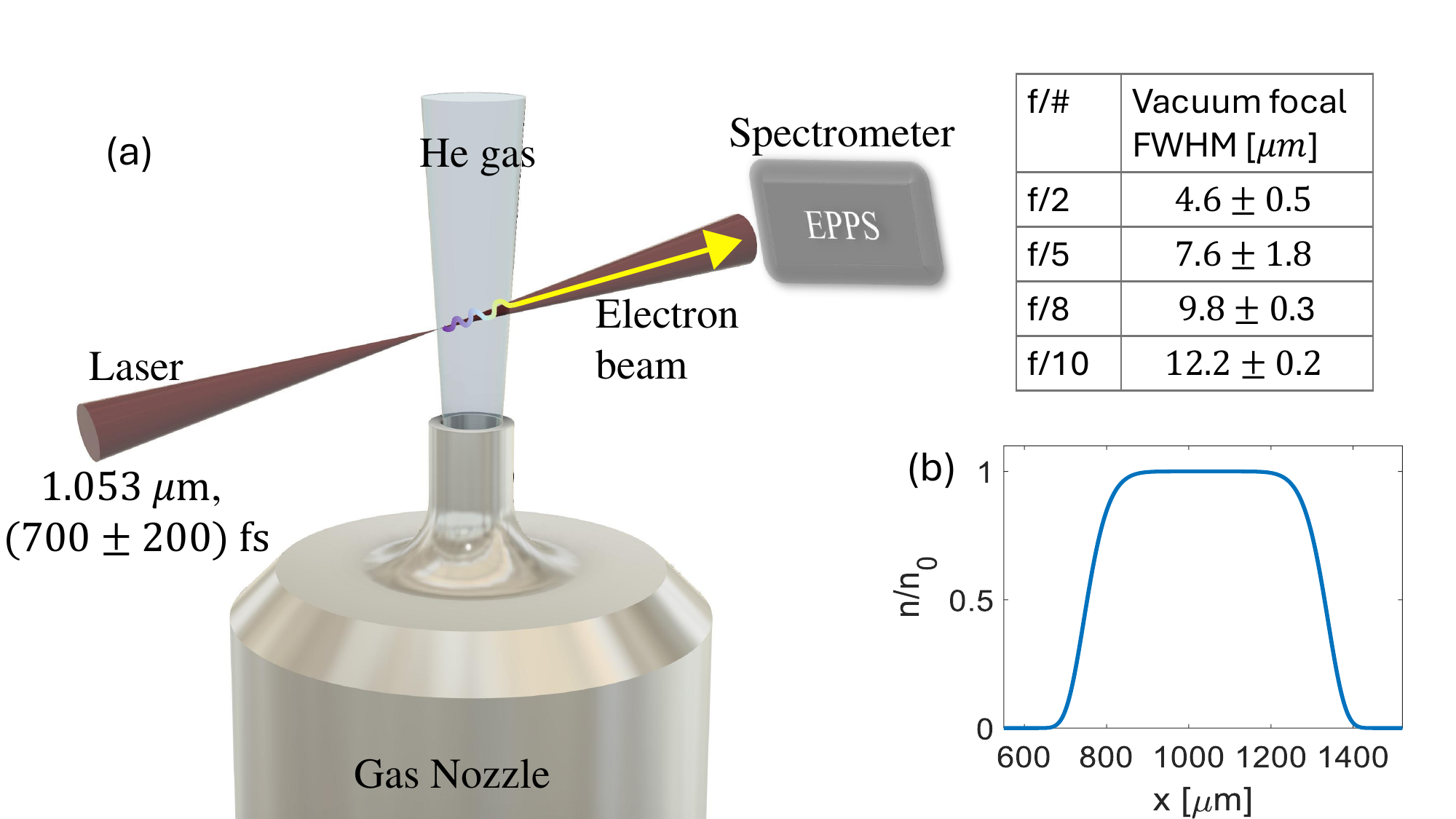}
		\caption{(a) Experimental setup. The table on the right shows $f$-number for four different apodizers and the corresponding in-vacuum laser FWHM. (b) The longitudinal density profile in simulations. $n_0$ is the peak plasma density, which is $0.02 ~n_c$.}
		\label{setup}
		\end{center}
\end{figure}
The experiment was performed at the University of Rochester Laboratory for Laser Energetics using the OMEGA EP laser facility.
Fig.\ \ref{setup}(a) shows the schematic of the setup.
A laser pulse with a central wavelength of $1.053~\mu\rm{m}$, duration of $(700\pm 200) \;$fs was focused by an $f/2$ equivalent off-axis parabolic mirror to the edge of a Mach~2~--~$2$~mm diameter supersonic helium gas jet target.
A wide range of peak plasma densities~($n_e$) from $0.008~n_c$ to $0.06~n_c$, where $n_c$ is the critical density, were examined and controlled by adjusting the backing pressure of the He gas.
The default near-field beam profile is square.
To change the effective $f$-number of the beam, circular apodizers with different diameters were used so that the full-width at half-maximum (FWHM) focal spot size in vacuum varied from $(4.6 \pm 0.5) ~\mu\rm{m}$ to $(12.2 \pm 0.2) ~\mu\rm{m}$. 
The on-shot wavefront was measured and the vacuum focal plane was reconstructed.
However, the increase of the beam size led to a reduction of the maximum possible on-target energy, hence limiting the obtainable $a_0$ for large focal spots.
The range of peak vacuum intensities was $(0.15$ -- $9.0)\times 10^{20}~\rm{Wcm}^{-2}$, corresponding to an $a_0$ range of $3.4$ to $27$.
The electron energy distributions were measured along the laser axis using a magnetic electron-positron-proton particle spectrometer (EPPS) \cite{chen2008} with an energy coverage from 1 – 150 MeV and the energy uncertainty range from $2\%$ at low energy end to up to $30\%$ at the high energy end \cite{von2021dispersion}.

\section{Particle-in-cell modeling}
\label{PIC}
To model the interaction, 2D PIC simulations were performed using the OSIRIS 4.0 code. 
A $[750 \; \mu\rm{m}\times 200 \; \mu\rm{m}]$ moving window with a resolution of 50 cells per $\lambda$ in longitudinal ($x$) direction and 35 cells per $\lambda$ in transverse ($y$) direction moving in the speed of light was utilized.
Each cell has 4 macroparticle electrons and 4 macroparticles representing fully ionized mobile helium ions.
An open boundary condition was applied to both dimensions.
To approximate the density distribution of a gas jet nozzle, the initial plasma density has a super-Gaussian profile in the $x$-direction, with a $450 \; \mu\rm{m}$ flat top area and two $150 \; \mu\rm{m}$ ramping zones connecting to the vacuum as shown in Fig.\ \ref{setup}(b), and a uniform distribution in the $y$-direction.
The maximum plasma density $n_0$ along the laser propagation axis was $0.01~n_{c}$, where $n_{c}=m_e \epsilon_0 \omega_0^2 / e^2$ is the critical plasma density and $\omega_0$ is the laser frequency.
The laser pulse was linearly polarized in $y$-direction with $a_0=3.5$, a wavelength of $1.053$~nm and a pulse duration of $\tau = 700$~fs.
It was launched from the vacuum region and focused at $n_e=0.95n_0$.
The initial laser electric field has a Gaussian spatial profile and a temporal form of $E=E_0 \sin(\pi t/\tau)$, where $\tau$ is the pulse duration.
Three different focal spot sizes with FWHM~$=~5 \; \mu\rm{m},~ 8 \; \mu\rm{m,}~ and ~16 \; \mu\rm{m}$ were examined.
And since the energy is changing for different focusing conditions, another set of simulations using a constant laser energy of 8.3~J, which is set based on the energy of the $8 \; \mu\rm{m}$ beam ($a_0=3.5$), were also performed.

\section{Results}
\label{experimental_results}
\begin{figure}
		\begin{center}
		\includegraphics[width= .7\linewidth]{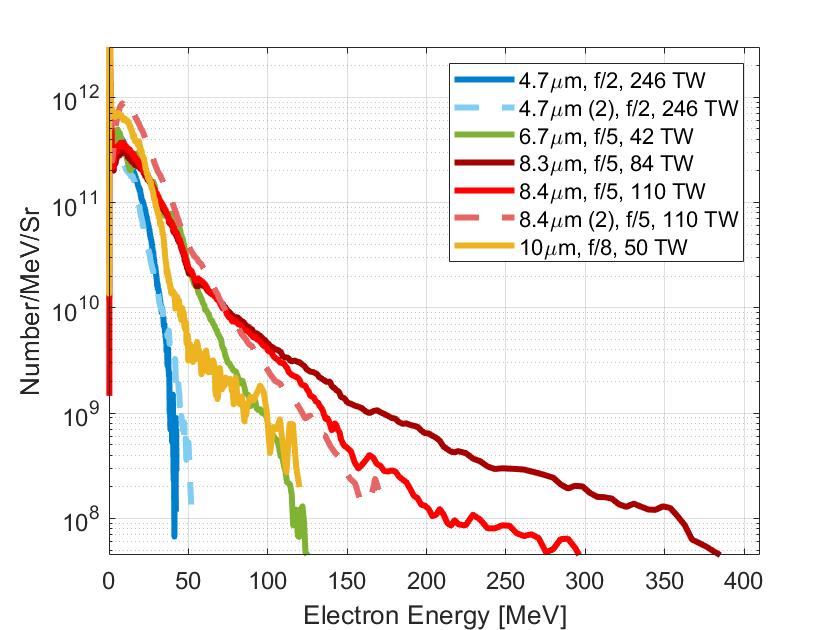}
		\caption{Experimental electron spectra for different laser focal spot sizes with similar peak plasma density of $(0.016\pm 0.004) ~ n_c$. 
        The light blue curve, labeled as \(4.7~\mu\rm{m}~(2)\), is a repeat shot of the dark blue shot. The light purple curve, labeled as \(8.4~\mu\rm{m}~(2)\), is a repeat shot of the dark purple shot. The laser in-vacuum focal spot size, f-number of the apodizer, and the laser power are shown in the legend. The laser \(a_0\) varies for different laser focusing conditions. From small beam size to large beam size, the laser \(a_0\) = 26 (FWHM = \(4.7~\mu\rm{m}\)), 7.8 (FWHM = \(6.7~\mu\rm{m}\)), 9 (FWHM = \(8.3~\mu\rm{m}\)), 10.3 (FWHM = \(8.4~\mu\rm{m}\)) and 5.7 (FWHM = \(10~\mu\rm{m}\)).}
		\label{exp_spectra}
		\end{center}
\end{figure}

\begin{figure}
	\centering
	\includegraphics[width= .7 \linewidth]{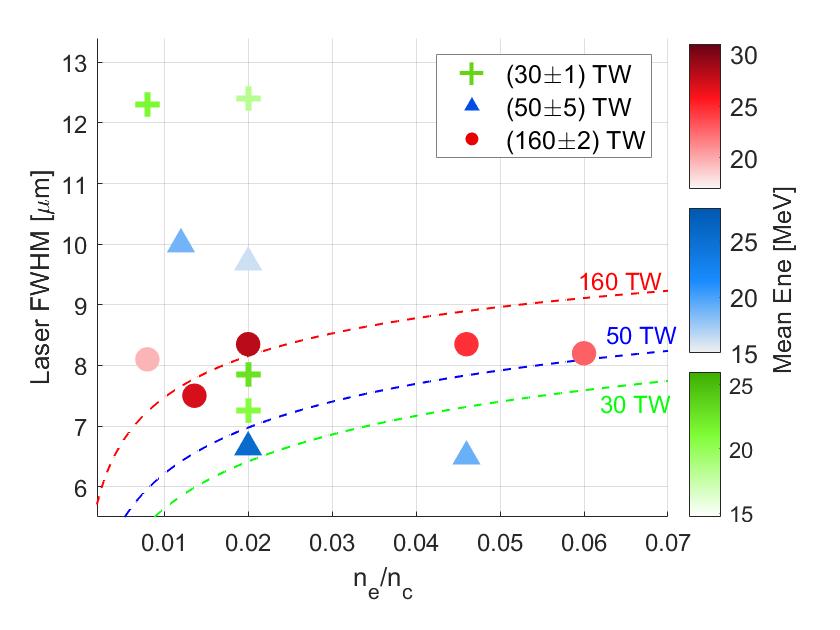}
	\caption{
    The scattered markers show the experimental data in the [ laser FWHM -- plasma density ] domain. Green plus markers are obtained using laser power of \( (30\pm1) \) TW, blue triangles are obtained using laser power of \( (50\pm 5) \) TW and red dots are obtained using laser power of \( (160\pm 2) \) TW. The color darkness indicates the electron mean energy, with darker color representing higher energy.
     \\
    The dashed curves show the theoretical prediction of the optimal conditions for electron acceleration, which is calculated based on the assumption of electron transverse displacement matching with laser beam size. The green, blue and red color correspond to laser power of $30 \; \rm{TW}$, $50 \; \rm{TW}$, and $160 \; \rm{TW}$ respectively.} 
	\label{matching_condition}
\end{figure}

\begin{figure}
	\centering
	\includegraphics[width=.9\linewidth]{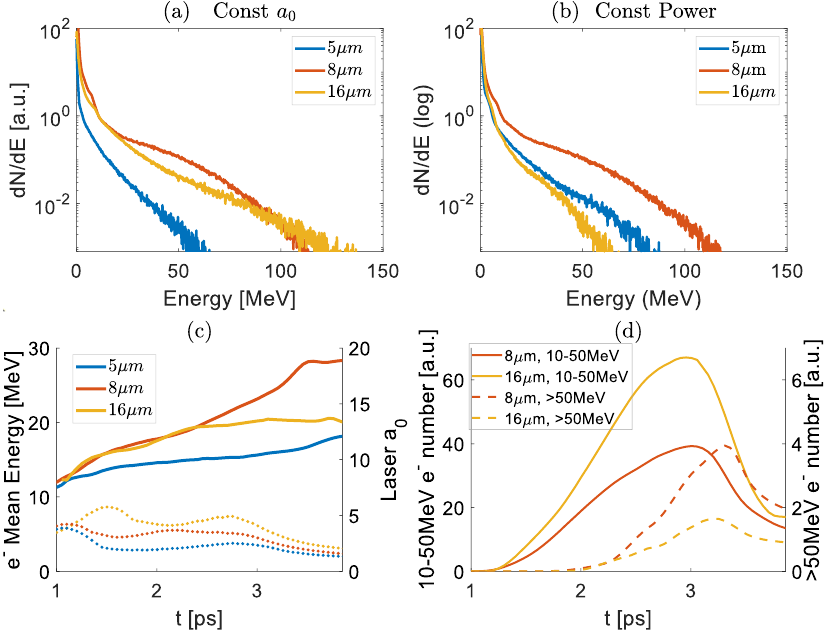}
	\caption{Simulated electron energy spectra for different initial focal spot sizes shown at $t=3.8$~ps using (a) constant laser $a_0$ and (b) constant laser power. (c) Temporal variation of laser $a_0$ (dashed lines) and the corresponding mean energy of electrons above cutoff energy of $10$~MeV (solid lines) from the same simulation sets of the panel (a). (d) Comparison of temporal variation of accelerated electron number for $8\; \mu\rm{m}$ and $16\; \mu\rm{m}$ beams from constant laser $a_0$ simulations. The solid lines represent electrons in the energy range of $10$--$50$~MeV and the dashed lines represent high-energy electrons $>50$~MeV.} 
	\label{fwhm_energy}
\end{figure} 

The highest electron energy and highest mean energy under the experimental conditions were achieved using a beam with moderate focal spot size. 
Fig.~\ref{exp_spectra} shows seven example experimental electron energy spectra from different laser focal spot sizes at a plasma density of $(0.016\pm 0.004)~n_c$.
The apodization of the beam to create the different focal spot sizes means the pulse energy was restricted for the large focal spot sizes, resulting in a wide range of $a_0 = 5.7$ -- $26$ for data shown in Fig.~\ref{exp_spectra}.
For a laser focal spot of $\rm{FWHM}=8.3 \; \mu\rm{m}$, a bi-Maxwellian-like electron energy distribution extending to $\sim 400$MeV was observed, which exceeds 120 times the
ponderomotive energy \(U_p=m_ec^2(\gamma -1)\), where \(\gamma=\sqrt{1+a_0^2/2}\).
Spectra with high energy tails exceeding 50~MeV are obtained for laser focal spot size within the range of $6.7 \; \mu\rm{m}$ to $10 \; \mu\rm{m}$.
The significant energy increase upon using a moderate-sized beam suggests there is an optimal laser focusing condition.
Due to constraints in experimental time, only two shots -- one with a laser beam size of \(4.7~\mu\rm{m}\) (dark blue curve in figure \ref{exp_spectra}) and another with \(8.4~\mu\rm{m}\) (bright red curve in figure \ref{exp_spectra}) -- were repeated to check the reproducibility.
Both repeated shots show minor shot-to-shot variations.
The \(4.7~\mu\rm{m}\) shot and its repeat shot almost overlap, with mean electron energies of 16.6 MeV and 16.8 MeV, respectively.
The repeat of the \(8.4~\mu\rm{m}\) shot shows a slightly larger gap.
The mean energies for the \(8.4~\mu\rm{m}\) shot and its repeat shot are 28~MeV and 23~MeV respectively.

A wider range of parameter space -- density, focal spot size, laser power -- was examined in experiments and more data were collected to better extract the energy variation trends.
A summary of the results is shown using scattered markers in Fig.~\ref{matching_condition}.
The data is divided into similar laser power, illustrated by the different colors -- \( (30\pm1) \)~TW (green), \( (50\pm 5)\)~TW (blue) and \( (160\pm 2)\)~TW (red) -- and the darkness of the color indicates the mean electron energy.
The mean electron energy was calculated for electrons above 10~MeV. 
However, due to the fact that changing the laser focal spot will inevitably change other parameters, such as laser intensity for a fixed laser energy or power, it is difficult to directly separate the individual effects of the laser focusing geometry. 
Therefore, to understand the complex relationship between the laser focusing geometry and the electron acceleration, we perform two sets of 2D PIC simulations: one with constant laser power and another with constant peak intensity.
This allows us to explicitly analyze the electron collective behavior and single particle dynamics.
In simulations, the final spectra of escaping electrons are diagnosed outside of the plasma, as the laser entirely leaves plasma at $t=3.8$~ps.
The results of simulations using a constant $a_0=3.5$ (corresponding to power of 5 TW, 12 TW, 47 TW for laser FWHM~$=~5 \; \mu\rm{m},~ 8 \; \mu\rm{m,}~ and ~16 \; \mu\rm{m}$) or constant laser power of 12 TW (corresponding to $a_0$ of 5.6, 3.5, 2 for laser FWHM~$=~5 \; \mu\rm{m},~ 8 \; \mu\rm{m,}~ and ~16 \; \mu\rm{m}$) are shown in Fig.~\ref{fwhm_energy} (a) and (b) respectively.
Enhanced acceleration is obtained using an $8 \; \mu\rm{m}$ beam in both cases and the improvement of electron number is more dramatic in Fig.~\ref{fwhm_energy} (b).
In addition, due to a low peak laser intensity, the mean electron energy of a $16\; \mu\rm{m}$ beam is lower than that of a $5\; \mu\rm{m}$ beam in the constant power simulation.


\begin{figure}
	\centering
	\includegraphics[width=1\linewidth]{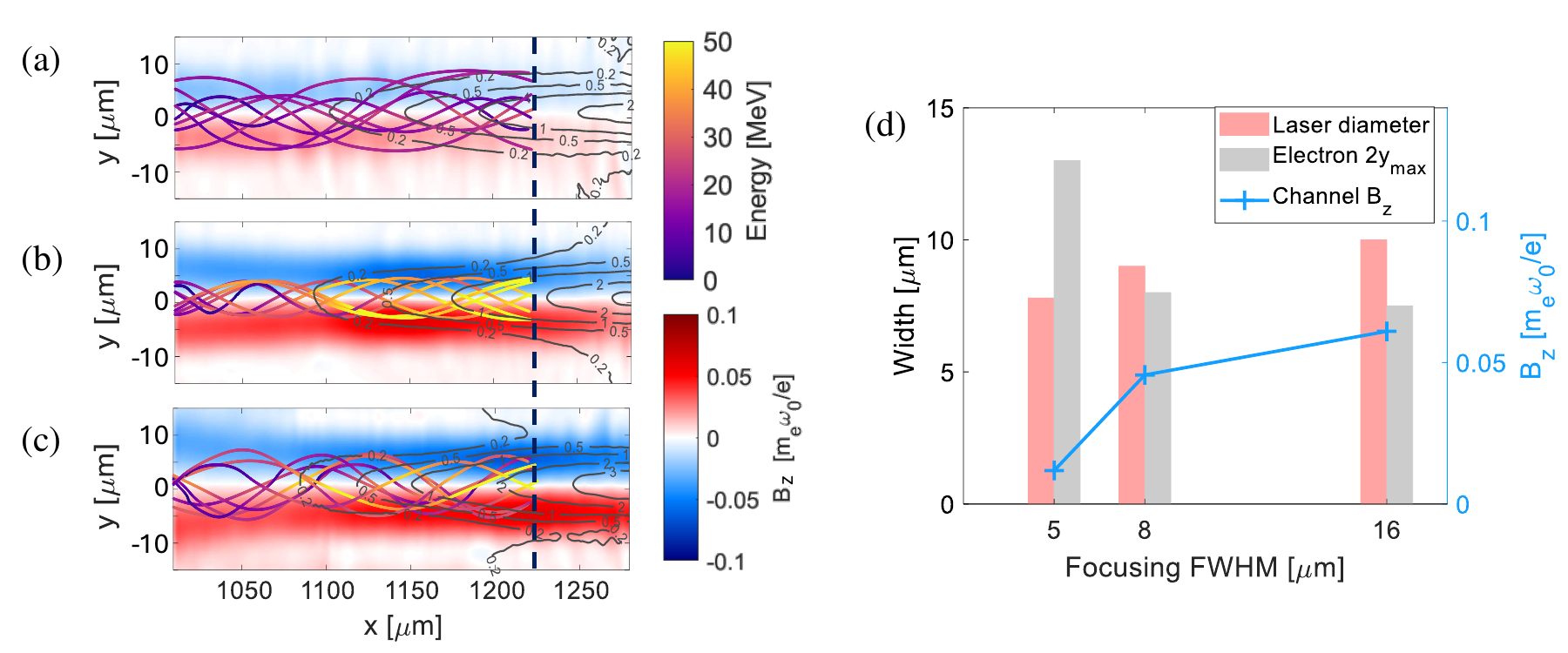}
	\caption{Electron trajectories overlapping on the channel $B_z$ fields at $t=2.86 \; \rm{ps}$ for laser beam size of (a) $5 \; \mu\rm{m}$, (b) $8 \; \mu\rm{m}$ and (d) $16 \; \mu\rm{m}$. The color of the electron trajectories represents electron energy. The black contours show the laser normalized field strength \(a\) value. (d) shows the laser beam diameter, the electron maximum transverse distribution range \(2y_{max}\), and the amplitude of channel $B_z$ fields at the position of $x=1225 \; \mu\rm{m}$. } 
	\label{field_tracking}
\end{figure} 
The 2D simulations provide insight into the influence of laser focusing on the laser field evolution, which is directly related to the channel formation and electron energy gain threshold.
Fig.~\ref{fwhm_energy} (c) shows the temporal evolution of the laser $a_0$ (dashed lines) and the corresponding electron mean energy (solid lines) for three focusing geometries.
The center of the laser pulse arrives at the initialized vacuum focal plane of $x=800 \; \mu\rm{m}$ at $t=1.1$~ps.
Then it propagates in the peak density region until $t=3$~ps, and finally the entire laser leaves the plasma at $t=3.8$~ps.
In the density up-ramp region, a beam initially with the smallest spot size
reaches peak intensity faster than a large beam.
As shown in Fig.~\ref{fwhm_energy} (c), the peak $a_0$ for a $5\; \mu\rm{m}$ beam appears around $1.1$~ps, corresponding to $10 \; \mu\rm{m}$ before the vacuum focus position in the spatial domain, while a $16\; \mu\rm{m}$ beam reaches its peak intensity at $t=1.5$~ps, which is roughly $100\; \mu\rm{m}$ after the vacuum focus point.
The $5\; \mu\rm{m}$ beam reaches its maximum $a_0$ prior to the vacuum focal plane and the small focal spot has a high transverse ponderomotive force resulting in a narrow, highly cavitated channel. 
Hence fewer electrons are available within the channel to be accelerated, leading to weaker currents and consequently weak self-generated magnetic fields, as shown in Fig.~\ref{field_tracking} (a).

After passing the focal plane, a small beam also rapidly defocuses and the $a_0$ is relatively low for the remaining interaction period.
The low laser intensity limits the maximum energy that electrons could gain.
In contrast to the tight focusing geometry, we see a more stable $a_0$ for an $8\; \mu\rm{m}$ beam.
After passing the vacuum focal plane, the laser defocuses to $a_0= 3$ at $t=1.6$~ps, then self-focuses to $a_0=3.5$ at $t=1.8$~ps.
The $a_0$ is sustained until it moves to the density down-ramp area. This allows the electron mean energy to increase with a relatively constant slope.

As the beam diameter increased to $16 \; \mu\rm{m}$ (corresponding to a higher power), the pulse focuses to highest $a_0 = 6$ at $t=1.5$~ps.
The laser intensity is better sustained along the entire interaction length compared to the smaller beams, with relatively large intensity fluctuations.
However, the significant intensity enhancement does not appear to improve the mean electron energy gain curve.
The temporal variation of the accelerated electron number is shown in Fig.~\ref{fwhm_energy} (d), for both electrons in the $10-50$~MeV and $>50$~MeV ranges. 
Before the laser reaches the density down-ramp region (at $t=3$~ps), the $16 \; \mu\rm{m}$ beam drives almost twice the number of low-energy ($10-50$~MeV) electrons compared to a $8 \; \mu\rm{m}$ beam.
However, fewer electrons are able to gain energy above $50$~MeV for the $16 \; \mu \rm{m}$ beam.
From 3~ps to 3.5~ps, a large number of electrons 
lose energy and gradually leave behind the laser. 
A sheath field is formed at the density down-ramp and channel exit due to the charge separation as the electron beam moves out from plasma to vacuum.
This causes the reduction in electron number, particularly for the lower energy electrons which could lose all the energy before exiting the sheath field.

 \begin{figure*}
	\centering
	\includegraphics[width=1\linewidth]{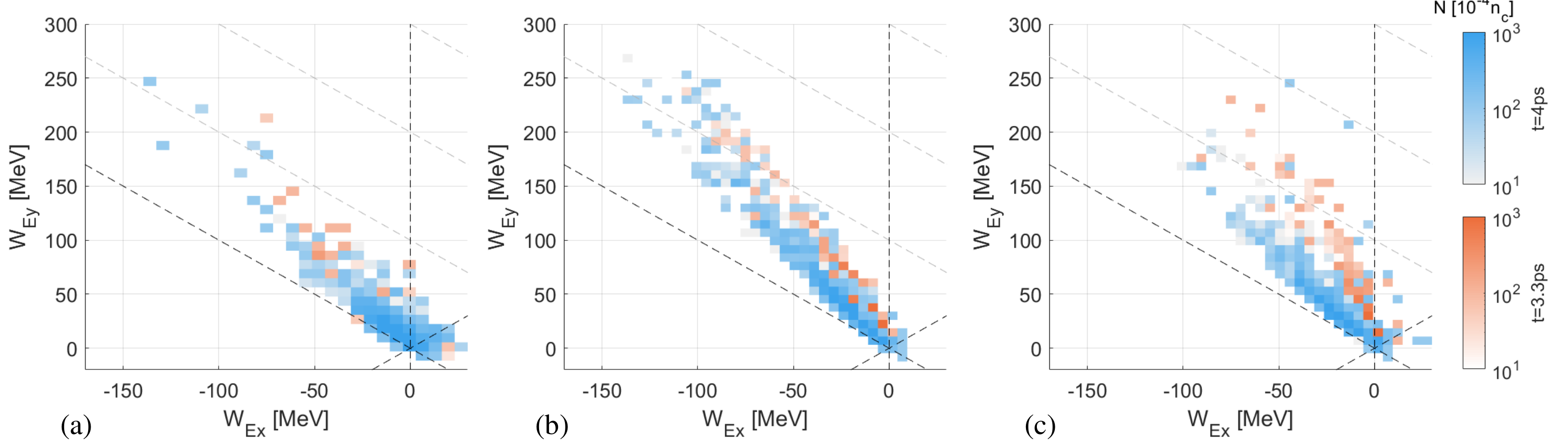}
	\caption{Electron distribution in the [$W_{E_x}-W_{E_y}$] energy space before (orange markers) and after (blue markers) the sheath field near down ramp region for laser focal spot size of $5 \; \mu\rm{m}$ (a), $8 \; \mu\rm{m}$ (b) and $16 \; \mu\rm{m}$ (c). The number of electrons is in the logarithm scale. The diagonal dotted lines are electrons with constant total energies ($W_{E_x}+W_{E_y}$), which are equal to the y-intercept.} 
	\label{work_FWHM}
\end{figure*}

The focusing geometry also implies laser longitudinal fields.  
Previous research has demonstrated that the transverse $E_y$ field does positive work for DLA electrons, whereas the longitudinal $E_x$ laser field tends to decelerate electrons \cite{arefiev2016}.
The ratio of laser transverse and longitudinal electric field is observed to be higher for the larger focal spots ($|E_y|/|E_x|_{5\mu\rm{m}}:|E_y|/|E_x|_{8\mu\rm{m}}=$ 19~:~26 at 2.2~ps).
Therefore, compared to the smallest focal spot, the longitudinal electric field of larger focal spots is expected to do relatively less negative work.

To investigate the effect of the longitudinal laser field 
and sheath field for each case, the energy contributions from different field components are calculated by performing particle tracking.
Seven hundred macro particles were randomly selected from electrons that move along with the laser and eventually out of the plasma.
They were tracked from the start of the simulations to the time when the laser completely exits the plasma.
The positions, momentum, and the exerted electric fields of the tagged electrons were diagnosed at every time interval of 25~fs. 
The work done by longitudinal electric field $E_x$ and transverse electric field $E_y$ are calculated by time integrals of $W_{E_x}=-\int_{0}^{t} |e| E_x\cdot v_x \, dt'$ and $W_{E_y}=-\int_{0}^{t} |e| E_y\cdot v_y \,dt'$, where $v_x$ and $v_y$ are electron longitudinal and transverse velocities respectively \cite{Gahn1999}.

Fig.~\ref{work_FWHM} shows the distribution of the tracked electrons in the $[W_{E_x}-W_{E_y}]$ energy space before ($t=3.3$~ps, orange markers) and after ($t=4$~ps, blue markers) the sheath field region for the three focusing geometries.
The electrons generated by a $5 \; \mu\rm{m}$ focal spot are more tilted towards $W_{E_x}$ axis in Fig.~\ref{work_FWHM} (a), indicating a greater proportion of negative work was done by the longitudinal laser electric field.
Also, the blue markers almost overlap with the orange ones, implying that the electron distribution is minimally affected by the sheath field.
For the moderate and large beams, the majority of pre-sheath field (orange) electrons are scattered around a direction of  $W_{E_y}/W_{E_x}\sim 3/1$ at $t=3.3$~ps, corresponding to the time of the high-energy electron number reaches peaks in Fig.~\ref{fwhm_energy} (d).
However, the difference is that fewer orange electrons exceed 50~MeV for a $16 \; \mu\rm{m}$ beam as shown in Fig.~\ref{work_FWHM}(c), as previously discussed.
Secondly, a larger number of electrons are driven out of plasma by a large focal spot (higher power pulse), hence a stronger sheath field grows in the down-ramp of the plasma, resulting in a greater impact on electron distribution.
The blue markers shift leftward along the negative $W_{E_x}$ direction and there is a wider gap between orange and blue electrons for the largest focal spot, suggesting more energy loss in the sheath field and a lower final energy compared to the mid-sized beam.


The particle tracking also reveals more details of the individual electron temporal and spatial dynamics.
Fig.~\ref{field_tracking} depicts the trajectories of eight typical energetic electrons overlaid on the instantaneous channel azimuthal magnetic fields $B_z$ (Fig.~\ref{field_tracking}(a) - (c)) at the time of electrons arriving at $x=1222 \; \mu\rm{m}$ for different laser focusing conditions.
And the black contours outline the laser normalized field strength \(a\).
Fig.~\ref{field_tracking}(b) clearly shows that electrons gain much higher energy with a moderate-sized driving pulse.
To better illustrate the relative scale of electron oscillation to the area that the laser field extends transversely, the maximum electron transverse position ($2y_{max}$) and laser beam diameter (\(2r\)) at $x=1222 \; \mu\rm{m}$ are presented using gray and red histograms in Fig.~\ref{field_tracking}(d). 
The peak channel $B_z$ field amplitude is also plotted in Fig.~\ref{field_tracking}(d).
As the channel expands with increased laser focal spot size, more electrons may be accommodated within the laser field and accelerated. 
As a consequence, stronger currents flow in the channel, forming a  stronger $B_z$ field. 
This self-generated quasi-static magnetic field assists the acceleration process by confining the electron radial excursion and deflecting the electrons in the forward direction \cite{Gong2020}.
Therefore, electrons are confined more tightly within the wider channel because of the stronger magnetic field, as shown by the gray histogram gradually becoming shorter than the red in Fig.\ \ref{field_tracking}(d). 
The ratio $r /\ y_{max}$ is 0.6, 1.1, and 1.4 for laser focal FWHM of $5 \; \mu\rm{m}$, $8 \; \mu\rm{m}$ and $16 \; \mu\rm{m}$ respectively.
Therefore, for a $5 \; \mu\rm{m}$ beam, the magnetic field is too weak to confine the electron in a way to efficiently allow DLA at $y_{max}$; for an $8 \; \mu\rm{m}$ beam, the electron transverse displacement is well-matched to the laser beam width; while for a $16 \; \mu\rm{m}$ beam, electrons are over-confined to a column narrower than the laser size.


Given that the maximum electron oscillation matches the laser beam size in the optimal scenario, a theoretical model can be developed to predict the optimal focusing geometry to a broader range of laboratory conditions.
In the laser channel, quasi-static electric and magnetic fields are generated and the laser beam size is approximately equal to the channel width. 
Assuming that the ponderomotive force is balanced by the force of the charge separation in a stable ion channel, the channel width ($\textit{w}$) can be estimated as approximately twice the laser spot radius ($r$) \cite{davis2005, lu2007}, so that $\textit{w}\approx 2r =2\cdot2\sqrt{a_0'}\cdot c/\omega_p$, where $a_0'$ is the normalized laser field strength when the laser is in the channel and $\omega_p$ is the plasma frequency. 
Since the laser power $P_L\propto a_0'^2r^2\sim a_0^2r_0^2$, the laser spot size can be rewritten as 
\begin{equation}
    r=w/2=[(2c/\omega_p)^2\cdot a_0r_0]^{1/3},
    \label{equation_r}
\end{equation}
where $r_0$ is the laser spot radius at the focal plane.
In ref.~\cite{Babjak2023, arefiev2014}, the maximum resonant electron transverse excursion from the channel axis is calculated using 
\begin{equation}
    y_{max}=(2c/\omega_p)\sqrt{\Big(\frac{a_0}{\epsilon}\cdot \frac{\omega_p}{\omega_0}\Big)^{2/3}-1},
    \label{equation_ymax}
\end{equation}
where $\epsilon$ is a parameter depending on the initial conditions. Here we take $\epsilon = 0.2$ for an ideal acceleration condition \cite{Babjak2023}.
Equating Eq.~(\ref{equation_r}) and Eq.~(\ref{equation_ymax}), 
\[
 (a_0r_0)^{1/3} =  (2c/\omega_p)^{1/3}\sqrt{\Big(\frac{a_0}{\epsilon}\cdot \frac{\omega_p}{\omega_0}\Big)^{2/3}-1}
\]
gives a condition where the electron transverse oscillation amplitude matches the laser spot size.

The results are plotted in Fig.~\ref{matching_condition} using dashed curves with green, blue, and red color showing laser power of $30 \; \rm{TW}, 50 \; \rm{TW}$ and $160 \; \rm{TW}$ respectively. 
For each certain curve, the laser power is constant.
Each position on this curve gives the optimal laser focusing FWHM for different plasma densities to reach maximum electron energy according to the simple model. 
Take the red curve for example, it has a fixed laser power of 160 TW, and the optimal laser FWHM for a plasma density of \(n_e = 0.02~n_c~\) is \(8.2~\mu\rm{m}\).
For each laser power, the optimal laser focal spot size increases with the plasma density.

To examine the validity of the simple theoretic model, the experimental data are plotted using scattered markers in the [ Laser~FWHM -- $n_e$ ] space in Fig.~\ref{matching_condition}.
The darkest red and blue markers, representing the highest mean energies measured, are located quite close to the red and blue dashed curves respectively.
This indicates that the optimized experimental parameter sets (laser power, laser focusing, and plasma density), which produced the electron beam with maximum mean energy, agree with the theoretical prediction of the ideal acceleration conditions.
For the low laser power of 30~TW, there is not enough experimental data showing a clear energy variation trend.  
Therefore,  at least for laser power above 30~TW, the theoretical calculation based on the assumption of electron oscillation matching with the laser beam size provides a good prediction of the optimal combination of laser and plasma parameters.

\section{Summary} \label{summary}
In conclusion, this work has experimentally demonstrated that optimizing the laser focusing geometry significantly enhances electron energy gain via DLA, and a maximum energy of \(\sim 400\)~MeV was observed using a hundred TW laser in experiments.
To focus the laser energy on the smallest possible focal spot to achieve the highest intensity is not always advantageous for DLA.
The optimal focusing geometry is achieved when the electron transverse oscillation amplitude matches the laser spot size in plasma, which is roughly the channel width. 
And based on the matching condition, a model is developed to find the optimal combination of laser power, focusing condition and plasma density in experiments.
Electrons from an optimal focal spot gain more energy than a tightly focused beam in bulk plasma and lose less energy than a very large beam in rarefied plasma and sheath field.
This demonstration and optimization of high-energy electron beams could lead the way towards ultrahigh-fluence x-ray delivered on a ps time scale \cite{rosmej2021}, which will dramatically enhance the radiographic capabilities of moderate relativistic intensity laser systems operating in high-energy-density research \cite{rusby2016}. 
Moreover, investigations into the channel formation and field evolution could assist fast ignition inertial confinement fusion scheme, where a strong collimating magnetic field is desired for guiding a large number of divergent ignition electrons generated in the coronal plasma \cite{tabak1994,scott2012}.

\section{Acknowledgements}
This work was supported by the Department of Energy National Nuclear Security Administration under Award Number DE-NA0004030.
The experiment was conducted at the Omega Laser Facility at the University of Rochester’s Laboratory for Laser Energetics with the beam time through the National Laser Users’ Facility (NLUF) program.
The authors would like to acknowledge the OSIRIS Consortium, consisting of UCLA and IST (Lisbon, Portugal) for providing access to the OSIRIS 4.0 framework. Work supported by NSF ACI-1339893.
The work of R.~Babjak and M.~Vranic was supported by Portuguese Foundation for Science and Technology (FCT) grants CEECIND/01906/2018, PTDC/FIS-PLA/3800/2021 DOI: 10.54499/PTDC/FIS-PLA/3800/2021 and FCT UI/BD/151560/2021.
The authors would like to acknowledge K.~McMillen for his gas jet plasma density calibration.
The work of K.~McMillen and J.~L.~Shaw was supported by the Department of Energy Office of Fusion Energy under Award Number DE-SC00215057 and by the Department of Energy National Nuclear Security Administration under Award Number DE-NA0004144.
The authors would also like to acknowledge G.~J.~Williams for his assistance with the calibration of the electron energy.

\section*{References}

\bibliographystyle{unsrt}
\bibliography{DLA}

\begin{thebibliography}{10}

\bibitem{strickland1985}
Donna Strickland and Gerard Mourou.
\newblock Compression of amplified chirped optical pulses.
\newblock {\em Optics Communications}, 55(6):447--449, 1985.

\bibitem{mourou1997}
G{\'e}rard~A Mourou, C~P Barty, and Michael~D Perry.
\newblock Ultrahigh-intensity laser: physics of the extreme on a tabletop.
\newblock 1997.

\bibitem{pukhov1999}
A~Pukhov, Z-M Sheng, and J~Meyer-ter Vehn.
\newblock Particle acceleration in relativistic laser channels.
\newblock {\em Physics of Plasmas}, 6(7):2847--2854, 1999.

\bibitem{Gahn1999}
C~Gahn, G~D Tsakiris, A~Pukhov, J~Meyer-ter Vehn, G~Pretzler, P~Thirolf, D~Habs, and K~J Witte.
\newblock Multi-mev electron beam generation by direct laser acceleration in high-density plasma channels.
\newblock {\em Physical Review Letters}, 83(23):4772, 1999.

\bibitem{meyer1999}
J~Meyer-ter Vehn and Zh~M Sheng.
\newblock On electron acceleration by intense laser pulses in the presence of a stochastic field.
\newblock {\em Physics of Plasmas}, 6(3):641--644, 1999.

\bibitem{willingale2006}
L~Willingale, SPD Mangles, PM~Nilson, RJ~Clarke, AE~Dangor, MC~Kaluza, S~Karsch, KL~Lancaster, WB~Mori, Z~Najmudin, et~al.
\newblock Collimated multi-mev ion beams from high-intensity laser interactions with underdense plasma.
\newblock {\em Physical Review Letters}, 96(24):245002, 2006.

\bibitem{Pomerantz2014}
Ishay Pomerantz, Eddie Mccary, Alexander~R Meadows, Alexey Arefiev, Aaron~C Bernstein, Clay Chester, Jose Cortez, Michael~E Donovan, Gilliss Dyer, Erhard~W Gaul, et~al.
\newblock Ultrashort pulsed neutron source.
\newblock {\em Physical Review Letters}, 113(18):184801, 2014.

\bibitem{Vranic2018}
Marija Vranic, Ondrej Klimo, Georg Korn, and Stefan Weber.
\newblock Multi-gev electron-positron beam generation from laser-electron scattering.
\newblock {\em Scientific Reports}, 8(1):1--11, 2018.

\bibitem{martinez2023creation}
Bertrand Martinez, Bernardo Barbosa, and Marija Vranic.
\newblock Creation and direct laser acceleration of positrons in a single stage.
\newblock {\em Physical Review Accelerators and Beams}, 26(1):011301, 2023.

\bibitem{Kneip2008}
S.~Kneip, S.~R. Nagel, C.~Bellei, N.~Bourgeois, A.~E. Dangor, A.~Gopal, R.~Heathcote, S.~P.~D. Mangles, J.~R. Marqu\`es, A.~Maksimchuk, P.~M. Nilson, K.~Ta Phuoc, S.~Reed, M.~Tzoufras, F.~S. Tsung, L.~Willingale, W.~B. Mori, A.~Rousse, K.~Krushelnick, and Z.~Najmudin.
\newblock Observation of synchrotron radiation from electrons accelerated in a petawatt-laser-generated plasma cavity.
\newblock {\em Physical Review Letters}, 100:105006, Mar 2008.

\bibitem{Ronga2021}
Maria~Grazia Ronga, Marco Cavallone, Annalisa Patriarca, Amelia~Maia Leite, Pierre Loap, Vincent Favaudon, Gilles Cr{\'e}hange, and Ludovic De~Marzi.
\newblock Back to the future: Very high-energy electrons (vhees) and their potential application in radiation therapy.
\newblock {\em Cancers}, 13(19):4942, 2021.

\bibitem{Morimoto2018}
Yuya Morimoto and Peter Baum.
\newblock Diffraction and microscopy with attosecond electron pulse trains.
\newblock {\em Nature Physics}, 14(3):252--256, 2018.

\bibitem{arefiev2016}
A~V Arefiev, V~N Khudik, A~P~L Robinson, G~Shvets, L~Willingale, and M~Schollmeier.
\newblock Beyond the ponderomotive limit: Direct laser acceleration of relativistic electrons in sub-critical plasmas.
\newblock {\em Physics of Plasmas}, 23(5):056704, 2016.

\bibitem{Kavin2023}
Kavin Tangtartharakul, Ilin Yeh, Hongmei Tang, Tao Wang, Louise Willingale, and Alexey Arefiev.
\newblock Mitigation of the detrimental role of the longitudinal laser electric field during direct laser acceleration of electrons.
\newblock {\em Bulletin of the American Physical Society}, 2022.

\bibitem{Babjak2023}
R.~Babjak, L.~Willingale, A.~Arefiev, and M.~Vranic.
\newblock Efficient direct laser acceleration by multi-petawatt lasers, 2023.

\bibitem{thomas2007}
A~G~R Thomas, Z~Najmudin, S~P~D Mangles, C~D Murphy, A~E Dangor, C~Kamperidis, K~L Lancaster, W~B Mori, P~A Norreys, W~Rozmus, et~al.
\newblock Effect of laser-focusing conditions on propagation and monoenergetic electron production in laser-wakefield accelerators.
\newblock {\em Physical Review Letters}, 98(9):095004, 2007.

\bibitem{Osiris}
Ricardo~A Fonseca, Luis~O Silva, Frank~S Tsung, Viktor~K Decyk, Wei Lu, Chuang Ren, Warren~B Mori, S~Deng, S~Lee, T~Katsouleas, et~al.
\newblock Osiris: A three-dimensional, fully relativistic particle in cell code for modeling plasma based accelerators.
\newblock In {\em International Conference on Computational Science}, pages 342--351. Springer, 2002.

\bibitem{Hemker2000}
Roy~Gerrit Hemker.
\newblock {Particle in cell modeling of plasma based accelerators in two-dimensions and three-dimensions}.
\newblock Thesis, 2000.

\bibitem{chen2008}
Hui Chen, Anthony~J Link, Roger Van~Maren, Pravesh~K Patel, Ronnie Shepherd, Scott~C Wilks, and Peter Beiersdorfer.
\newblock High performance compact magnetic spectrometers for energetic ion and electron measurement in ultraintense short pulse laser solid interactions.
\newblock {\em Review of Scientific Instruments}, 79(10):10E533, 2008.

\bibitem{von2021dispersion}
Jens von~der Linden, Jos{\'e} Ramos-M{\'e}ndez, Bruce Faddegon, Devan Massin, Gennady Fiksel, Joe~P Holder, Louise Willingale, Jonathan Peebles, Matthew~R Edwards, and Hui Chen.
\newblock Dispersion calibration for the national ignition facility electron--positron--proton spectrometers for intense laser matter interactions.
\newblock {\em Review of Scientific Instruments}, 92(3), 2021.

\bibitem{Gong2020}
Z~Gong, Felix Mackenroth, T~Wang, X~Q Yan, T~Toncian, A~V Arefiev, et~al.
\newblock Direct laser acceleration of electrons assisted by strong laser-driven azimuthal plasma magnetic fields.
\newblock {\em Physical Review E}, 102(1):013206, 2020.

\bibitem{davis2005}
J~Davis, GM~Petrov, and AL~Velikovich.
\newblock Dynamics of intense laser channel formation in an underdense plasma.
\newblock {\em Physics of Plasmas}, 12(12):123102, 2005.

\bibitem{lu2007}
Wei Lu, M~Tzoufras, C~Joshi, FS~Tsung, WB~Mori, J~Vieira, RA~Fonseca, and LO~Silva.
\newblock Generating multi-gev electron bunches using single stage laser wakefield acceleration in a 3d nonlinear regime.
\newblock {\em Physical Review Special Topics-Accelerators and Beams}, 10(6):061301, 2007.

\bibitem{arefiev2014}
Alexey~V Arefiev, Vladimir~N Khudik, and Marius Schollmeier.
\newblock Enhancement of laser-driven electron acceleration in an ion channel.
\newblock {\em Physics of Plasmas}, 21(3):033104, 2014.

\bibitem{rosmej2021}
ON~Rosmej, XF~Shen, A~Pukhov, L~Antonelli, F~Barbato, M~Gyrdymov, MM~G{\"u}nther, S~Z{\"a}hter, VS~Popov, NG~Borisenko, et~al.
\newblock Bright betatron radiation from direct-laser-accelerated electrons at moderate relativistic laser intensity.
\newblock {\em Matter and Radiation at Extremes}, 6(4):048401, 2021.

\bibitem{rusby2016}
DR~Rusby, CM~Brenner, Chris Armstrong, LA~Wilson, R~Clarke, A~Alejo, H~Ahmed, NMH Butler, D~Haddock, A~Higginson, et~al.
\newblock Pulsed x-ray imaging of high-density objects using a ten picosecond high-intensity laser driver.
\newblock In {\em Emerging Imaging and Sensing Technologies}, volume 9992, pages 61--68. SPIE, 2016.

\bibitem{tabak1994}
Max Tabak, James Hammer, Michael~E Glinsky, William~L Kruer, Scott~C Wilks, John Woodworth, E~Michael Campbell, Michael~D Perry, and Rodney~J Mason.
\newblock Ignition and high gain with ultrapowerful lasers.
\newblock {\em Physics of Plasmas}, 1(5):1626--1634, 1994.

\bibitem{scott2012}
RHH Scott, F~Perez, JJ~Santos, CP~Ridgers, JR~Davies, KL~Lancaster, SD~Baton, Ph~Nicolai, RMGM Trines, AR~Bell, et~al.
\newblock A study of fast electron energy transport in relativistically intense laser-plasma interactions with large density scalelengths.
\newblock {\em Physics of Plasmas}, 19(5):053104, 2012.

\end{thebibliography}

\end{document}